\documentclass[aps,prl,twocolumn,showpacs]{revtex4}

\usepackage[dvips]{graphicx}
\usepackage{amssymb}
\usepackage{amsmath}

\begin{document}

\title{Frictional dissipation of polymeric solids {\it vs} interfacial glass transition}

\author{Lionel Bureau}
 \email{bureau@insp.jussieu.fr}
\author{Christiane Caroli}
\author{Tristan Baumberger}

\affiliation{Institut des Nanosciences de Paris, UMR 7588 CNRS-Universit\'e Paris 6, 140 rue de Lourmel, 75015 Paris, France}

\date{\today}

\begin{abstract}
We present single contact friction experiments between a glassy polymer and smooth silica substrates grafted with alkylsilane layers of different coverage densities and morphologies.
This allows us to adjust 
the polymer/substrate interaction strength. We find that, when going from weak to strong interaction, the response of the interfacial junction where shear localizes evolves from that 
of a highly viscous threshold fluid
to that of a plastically deformed glassy solid. This we analyse as resulting from an interaction-induced ``interfacial glass transition'' helped by pressure.
\end{abstract}


\maketitle

The microscopic origin of solid friction has been the focus of a growing number of studies
 over the past decade. Parallel efforts have been made --- using atomic force microscopy (AFM) \cite{salmRev97,barrena,riedo,overney}, surface force apparatus (SFA) \cite{SFA1,SFA2,SFA3} and 
 sphere/flat tribometers \cite{chaudhury,nonamontons} --- 
 to investigate situations of single contact friction at the nanometer, micrometer, and hundred of micrometers scale.
Two main theoretical pictures have been proposed in order to account for phenomenologies observed for large scale (micrometer and above) contacts.

{\it (i)} The first one relates friction to the yield properties of a jammed confined layer which responds to shear as an {\it elastic-plastic disordered solid} \cite{persson}:
 dissipation is associated with the sudden flips of bistable molecular-sized ``shear transformation zones'', these elementary events
taking place in the interfacial ``junction'' of molecular thickness where strain localizes. 
Such a description, when extended to include thermal activation effects, has proven to account properly for the shear stress logarithmic
velocity dependence  observed for glassy polymers \cite{nonamontons}. It has also been invoked to explain 
the solid-like response of boundary lubricants confined down to molecular thicknesses \cite{lemaitre}.

{\it (ii)} The second picture is well suited to account for 
friction of rubbers \cite{chaudhury} on  smooth substrates, or sliding
of surfactant-bearing surfaces \cite{richetti}, {\it i.e.} interfaces which, even though they exhibit a static threshold, 
cannot be depicted as jammed solid junctions as mentioned above. Static friction results from the rupture of adhesive molecular bonds (or links) that can form between the solids. 
Kinetic friction is then the combined result of the non-trivial dynamics of this {\it depinning/repinning} process \cite{schallamach} and of the {\it viscous} contribution of the 
sheared interfacial layer \cite{urbakh,richetti,tbcc}. 

These two pictures have been proposed and used in separate contexts.
Recently, it has been hypothesized that, for a given couple of bulk solids, it should be possible to cross over from one situation to the other by tuning 
the corrugation of the surface interaction potentials \cite{tbcc}, {\it i.e.} the physico-chemical 
nature of the surfaces confining the interfacial junction.

In this Letter, we report clear experimental evidence of such a crossover. We perform single contact friction experiments between a glassy polymer and 
silicon substrates with different densities of strong pinning sites 
available for the polymer chains. We observe qualitative changes in the velocity dependence of the friction stress $\sigma (V)$, which, when analyzed,
 lead us to propose the following unified picture of dissipation at the interface between a polymeric solid and a smooth
substrate: as the pinning interaction strength increases, the interfacial junction undergoes a glass transition, while frictional dissipation commutes from a threshold fluid rheology to 
2D weak glass plasticity.

\begin{figure}[htbp]
\includegraphics[scale=0.6]{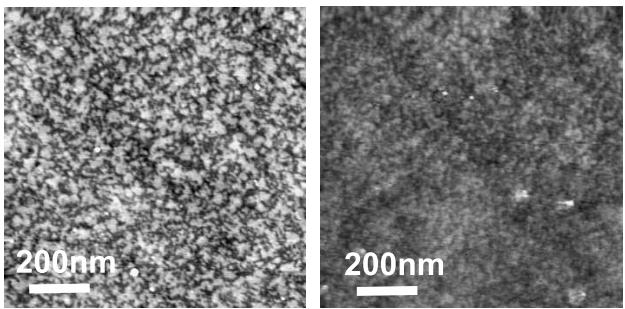}
\caption{AFM topographic images (tapping mode) of (a) OTS18 and (b) OTS25. Height scale is 26\AA~from black
(low) to white (high).}
\label{fig:afm}
\end{figure}

Experiments are performed on a homebuilt tribometer (see \cite{nonamontons} for a detailed description) in which a smooth lens of 
glassy poly(methylmethacrylate) (PMMA, $T_{g}\simeq110^{\circ}$C) is pressed against a flat silicon substrate. At constant normal load $N$, we
simultaneously monitor the contact area $A$ (measured optically through the polymer lens) and the tangential force $F$ while the subtrate is driven at constant 
velocity through a compliant load cell. We thus have access to the shear stress $\sigma=F/A$ as a function of the applied contact 
pressure $p=N/A$ (in the range 7---80 MPa) and the driving velocity $V$ (in the range 0.1---300 $\mu$m.s$^{-1}$). All measurements are made at 
$T=22^{\circ}$C, in a glovebox purged with dry argon. PMMA samples are prepared following a protocole \cite{nonamontons} which allows us to obtain 
lenses of millimetric radius of curvature, and of a root-mean-square roughness of $\sim$3~\AA~at their apex 
 (as measured by AFM over a 1 $\mu$m$^{2}$ scan). Substrates are silicon wafers covered by a nanometer-thick native oxide layer. They are cleaned in 
 a UV/O$_{3}$ chamber for 30 min, and subsequently exposed to a water saturated oxygen flux in order to prepare hydroxylated silica surfaces, {\it i.e.} exhibiting a high number
 of silanol (Si-OH) groups.  Three different types of methyl-terminated silane layers can be grafted on these surfaces: (i) a  layer of trimethylsilane
 (TMS) obtained from gaz-phase grafting of hexamethyldisilazane, (ii) a layer of octadecyltrichlorosilane (OTS) obtained by immersing a substrate for 5 min in a solution of OTS in 
 hexadecane/carbon tetrachloride at 18$^{\circ}$C (labeled OTS18 in the following), (iii) an OTS layer grafted in a solution maintained at 25$^{\circ}$C (labelled OTS25).
The first type of silane (TMS) does not self-assemble and yields a thin ($\lesssim$ 5\AA) disordered {\it sub}monolayer in which non-passivated Si-OH groups are still available. Changing the
grafting temperature of OTS has an impact on the layer morphology, as documented in extensive studies \cite{goldmann}. OTS18 has an average thickness $\sim 21$\AA 
(as measured by ellipsometry) 
and exhibits islands of high areal density of silane molecules, separated by regions of much lower coverage density or even bare substrate (Fig. \ref{fig:afm}a). On the other 
hand, OTS25 has an average thickness $\sim$15\AA~, but presents a more uniform coverage density (Fig. \ref{fig:afm}b). We therefore expect the density of available  
silanol groups to be low on OTS25, slightly higher on OTS18, and much higher on TMS coated wafers. The PMMA macromolecules can form hydrogen 
bonds, via their carbonyl groups, with these silanols, which thus act as pinning centers for the polymer chains \footnote{The work of adhesion, obtained from a JKR fit of the 
load/contact radius curves, is found to be much higher for PMMA/TMS than for PMMA/OTS25 (0.4$\pm$0.1 J.m$^{-2}$ $vs$ 0.08$\pm$0.01 J.m$^{-2}$), in agreement
with strong ``H-bond pinning'' of the polymer on TMS coated substrate. Differences between OTS18 and OTS25, not detectable from the estimated work of adhesion, are suggested by the distinct 
morphologies observed by AFM.}. 

\begin{figure}[htbp]
\includegraphics[scale=0.8]{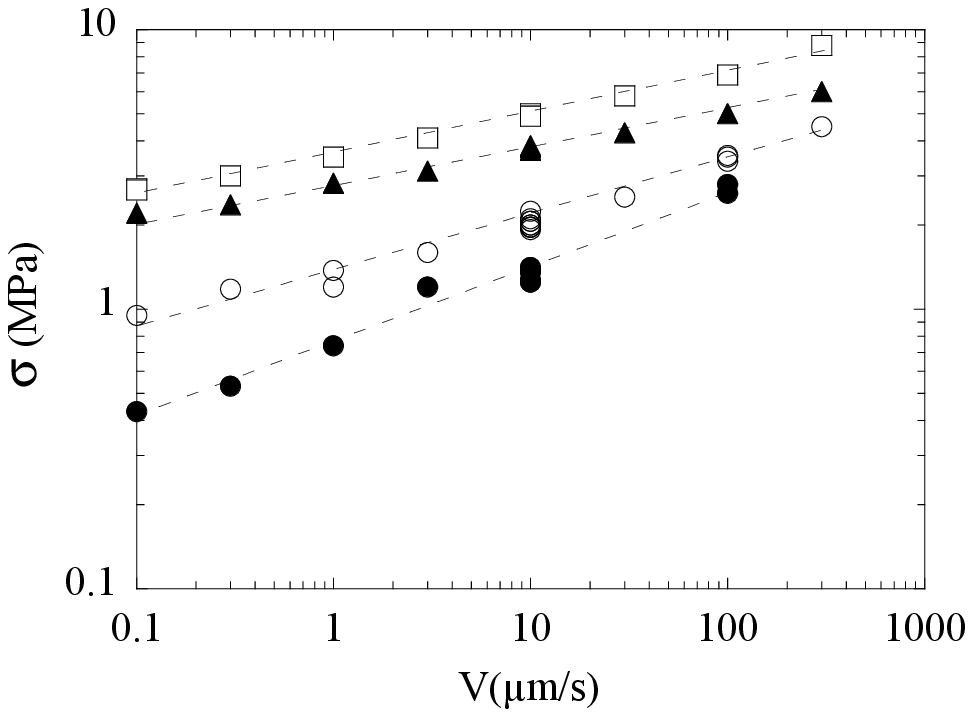}
\caption{Shear stress $vs$ sliding velocity for PMMA on OTS25. ($\bullet$) p=10 MPa; ($\circ$) p=19 MPa;
($\blacktriangle$) p=38 MPa; ($\square$) p=63 MPa.}
\label{fig:sigVOTS25}
\end{figure}

{\it Low pinning level} --- The velocity dependence of the shear stress, measured in steady sliding under various contact pressures, is shown on Fig. \ref{fig:sigVOTS25} for the OTS25 substrate.  

$\sigma$ is found to increase 
as $V^{\alpha}$, with $\alpha$ decreasing from 0.25 to 0.15 when $p$ grows from 10 to 60 MPa. The shear stress, in the range 0.5--10 MPa, is much lower than the
shear yield stress of PMMA ($\sim$70 MPa), and bulk dissipation in the polymer can thus be ruled out.  On the other hand, under the contact pressures used here, 
 the contribution to energy dissipation arising from molecular rearrangements in the compressed OTS layer is expected to be negligible \footnote{If we assume that, due
to loose packing of silane molecules in OTS25, a mechanism such as chain tilting could contribute to dissipation, we can estimate the resulting shear stress as follows: $\sigma A d=nE$, 
where $A$ is the contact area, $d$ the slid distance, $E$ the energy per molecule needed to induce tilting, and $n$ the number of OTS molecules swept during a displacement of $d$.
Taking $A\sim (100\, \mu m)^{2}$, $E\sim 0.3$ eV per molecule \cite{barrena}, and an area per molecule of $\sim 25$\AA$^{2}$ \cite{barrena}, we obtain $\sigma \sim 1$ kPa. }.

We therefore propose that the interfacial response, in this situation of weak polymer/substrate interaction, is governed, in the V-range explored, by the {\it viscous flow} of a liquidlike layer at the surface of the 
PMMA sample, the mobility of these surface chains being lower when pressure is higher. This is supported by the following facts. {\it (i)} It has been shown recently that the glass transition shift in 
PMMA thin films supported on OTS-coated wafers is 
consistent with the existence, at the film/substrate interface, of a nanometer-thick polymer layer of enhanced segmental mobility with respect to the bulk \cite{nealey}. {\it (ii)}
The $V^{\alpha}$ dependence of $\sigma$ is akin to the shear-thinning behavior of confined polymer melts \cite{yamada,leger}. 
 Moreover, assuming that the junction has a thickness $h\sim 1$ nm, we can make a tentative estimate of its viscosity  $\eta_{eff}=\sigma V/h$: at the lowest shear rate (100 s$^{-1}$), 
 we thus find that 
$\eta_{eff}$ grows from $5.10^{3}$ to $5.10^{4}$ Pa.s when $p$ inceases from 10 to 60 MPa. Such a range for $\eta_{eff}$ is consistent with the viscosities reported for 
 PMMA melts of very short chains at 20--30 K above their $T_{g}$ \cite{oconnor}. 
\begin{figure}[htbp]
\includegraphics[scale=1]{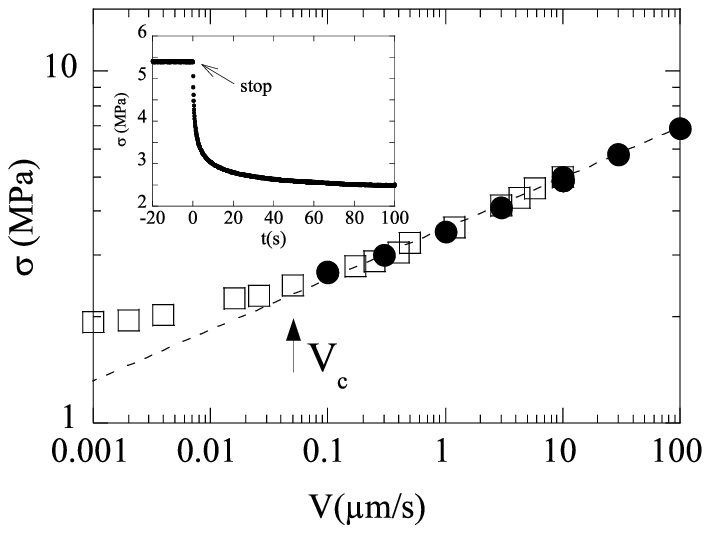}
\caption{$\sigma (V)$ measured in steady-state ($\bullet$) and in transient ($\square$), at $p=63$ MPa, for PMMA sliding on OTS25. Inset: stress relaxation $\sigma (t)$ following a stop
of the drive.}
\label{fig:crossover}
\end{figure}
 
We further investigate the shear behaviour of the junction by performing stress relaxation experiments as follows: starting from steady sliding at constant
velocity, we stop driving the tangential loading spring, and monitor the subsequent slow stress drop $\sigma (t)$, as illustrated
in the inset of Fig. \ref{fig:crossover}. We compute the instantaneous sliding velocity $\dot{x}$ from $\sigma(t)$, and 
thus have access to $\sigma (\dot{x})$ during the relaxation transient. As seen on Fig. \ref{fig:crossover}, two distinct regimes appear.

 {\it (i)} 
for $V\gtrsim V_{c}=50$--100~nm.s$^{-1}$ stress-velocity data obtained in transient and in steady-state collapse on the same curve, indicating that, in this regime, the
shear stress depends only on the instantaneous sliding velocity. 

{\it (ii)} for $V<V_{c}$, $\sigma (\dot{x})$ 
clearly deviates from the power law and the
shear stress tends towards a finite value at low velocities, which reveals the existence of a static friction threshold.

\begin{figure}[htbp]
\includegraphics[scale=1]{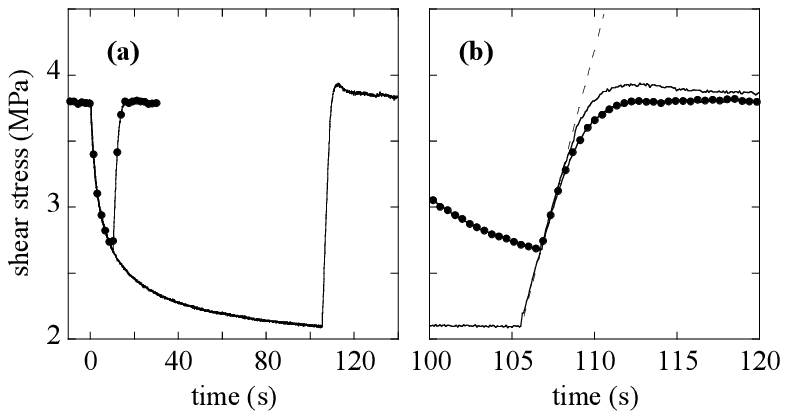}
\caption{PMMA on OTS25 at $p=38$ MPa. (a) Shear stress $vs$ time during the following sequence: steady-sliding at 10 $\mu$m.s$^{-1}$ ($t<0$), stop (10 and 100 seconds), and
reloading at 10 $\mu$m.s$^{-1}$ until steady-sliding. Just before reloading, the instantaneous sliding velocity is 70 nm.s$^{-1}$ for the 10s relaxation, and 2 nm.s$^{-1}$ for the 100s
relaxation. (b) Close-up view of the response upon reloading after 10s (full line labeled with $\bullet$) and 100s (full line) stops. 10s data have been horizontally shifted to have both loading phases 
coincide. The straight dotted line is a guide for the eye.}
\label{fig:fig4}
\end{figure}

We probe the build-up of the static threshold through the interfacial response upon reloading after a
stress relaxation of given duration. After short relaxations --- at 
the end of which
the instantaneous sliding velocity is still above the crossover velocity $V_c$ --- we observe a monotonous viscouslike transient before reaching the steady-state constant stress (Fig. \ref{fig:fig4}). 
On the contrary, for 
longer stress relaxations, with 
final sliding velocities below $V_{c}$, an elastic response followed by a stress overshoot is observable (Fig. \ref{fig:fig4}). From this we conclude that, for  $V<V_{c}$, pinning occurs at the interface.
 
These observations lead us to propose the following picture, 
in the spirit of Schallamach's model of rubber friction. Pinning of PMMA segments on the substrate
 is expected to become effective at sliding velocities $V<V_{c}= d/\tau$, where $d$ is an average capture radius
of the pinning sites, and $\tau$ a characteristic relaxation time of the surface polymer chains. In this low $V$ regime, the number of strong
bonds that can form between the polymer and the substrate decreases when the sliding velocity increases. This leads, on top of
the viscous one, to a velocity-weakening contribution to $\sigma (V)$, which, if strong enough,
results in a V-weakening regime. That no such regime is observed on OTS25 means that, if it exists, it lies below the accessible velocity range. For $V>V_{c}$, pinning becomes negligible, 
and dissipation 
is controlled solely by the viscosity of the unpinned liquid layer. 

\begin{figure}[tbp]
$$
\includegraphics[scale=0.8]{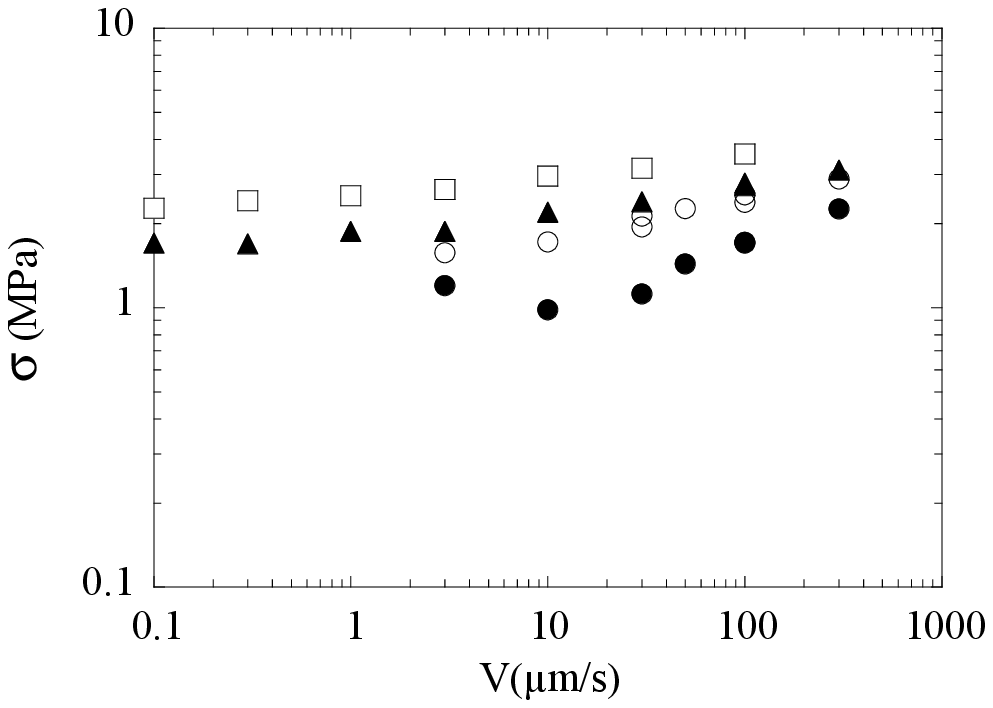}
$$
\caption{$\sigma (V)$ measured in steady-state for PMMA sliding on OTS18. ($\square$) $p=55$ MPa; ($\blacktriangle$) $p=30$ MPa; ($\circ$) $p=15$ MPa;
($\bullet$) $p$=7 MPa. Unstable sliding is observed for $V<3\, \mu$m.s$^{-1}$ at $p=7$ and $V\leq 1\, \mu$m.s$^{-1}$ at 15 MPa.}
\label{fig:sigVOTS18}
\end{figure}

{\it Intermediate pinning level} --- In the above description, the depinning threshold depends on the size of the pinning sites $d$, and on the relaxation time $\tau$. 
Changing $d$ or $\tau$ should then allow to shift $V_{c}$ up to values in the experimental velocity window. This we perform by using the OTS18 substrate, which presents larger
coverage defects, hence a larger $d$, than OTS25. Comparison of Fig. \ref{fig:sigVOTS25} and \ref{fig:sigVOTS18} shows that this results in marked qualitative differences for $\sigma (V)$.
Whereas, for OTS25, the power law behavior is observed for all contact pressures, for OTS18 at $p=7$ MPa, $\sigma(V)$ does exhibit a minimum at $V_{c}=10\, \mu$m.s$^{-1}$. 
The velocity-weakening regime below $V_{c}$ leads to a stick-slip instability that appears for $V<3\, \mu$m.s$^{-1}$. For $p=15$ MPa, stick-slip also occurs at $V\leq 1\, \mu$m.s$^{-1}$, indicative
here again of an underlying velocity-weakening. For $p \geq 30$ MPa, steady sliding is observed over the whole velocity range, but a flattening of the $\sigma (V)$ curve is still visible at 
its low-$V$ end. That is, increasing $p$ shifts $V_{c}$ towards lower values, which indicates that the higher the pressure, the larger $\tau$, or equivalently the lower the mobility of the
confined PMMA surface segments. 

\begin{figure}[tbp]
$$
\includegraphics[scale=0.8]{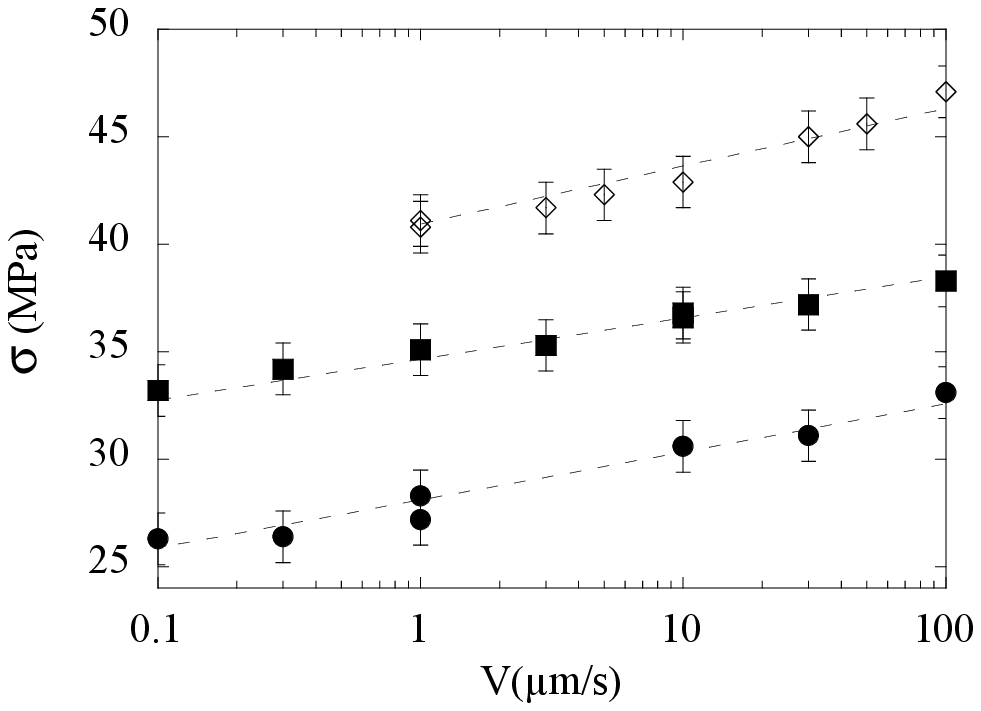}
$$
\caption{$\sigma (V)$ for PMMA sliding on TMS. ($\bullet$) $p=35$ MPa; ($\blacksquare$) $p=60$ MPa; ($\diamond$) $p=300$ MPa. Data at $p=300$ MPa were obtained from
friction experiments between a rough PMMA block and a smooth TMS coated substrate; the shear stress is taken as $\sigma=\mu p$, where $\mu$ is the measured friction coefficient and
$p$ is assumed to be equal to the hardness of PMMA (300 MPa).}
\label{fig:sigVTMS}
\end{figure}

{\it High pinning level} --- As previously reported \cite{nonamontons}, strong pinning is realized with the help of the TMS substrate. In this latter case, $\sigma$ is in the range 25--50 MPa, {\it i.e.} about
 one order of magnitude higher than on OTS25, and
displays a logarithmic increase with $V$, as shown on 
Fig. \ref{fig:sigVTMS}. Its log-slope depends only weakly on pressure. Such a $\ln (V)$ dependence of $\sigma$ has been shown to result
from thermally-assisted stress-induced structural rearrangements. These rearrangements involve zones of $\sim$nm$^{3}$ \footnote{From the log-slopes of $\sigma (V)$, we compute
an activation volume in the range 3--5 nm$^{3}$} localized in a nanometer-thick polymer layer, which, under low shear stresses, behaves as a solid {\it glassy} medium 
\cite{nonamontons,rugueuxlisse}. 

In conclusion, we have shown here that, when the strength of interfacial interactions is increased, the frictional rheology gradually evolves from that of a highly viscous threshold liquid
to that of a plastically deformed glassy medium. We therefore conclude that the nanometer-thick junction where shear localizes undergoes an ``interfacial glass transition''. That is, in a situation
where molecular mobility at the free PMMA surface is still liquid-like, confinement by a strongly corrugated potential, helped by pressure, is able to quench the interface into a jammed glassy state. 
This interpretation is fully consistent with the analysis of glass transition shifts in thin supported films in terms of the existence, at the surface of a glassy polymer, of
a layer in which chain mobility is all the higher as the interaction with the substrate is weak \cite{nealey}. It shows that sliding friction may be turned into a highly sensitive probe of polymer 
surface dynamics.

\begin{acknowledgments}
We are grateful to  M. Goldmann for valuable advice about self-assembled monolayer deposition, and to E. Lacaze and B. Gallas for their help with substrate characterization.

\end{acknowledgments}

\bibliography{billeplan2}

\end{document}